\documentclass [12pt]{article}

\begin{document}
\begin{center}
\textbf{COINCIDENCE OF THE UNIVERSE DESCRIPTION STEMMING FROM D-BRANES THEORY AND ENU MODEL}
\end{center}

\bigskip

\begin{center}
Jozef \v{S}ima and Miroslav S\'uken\'{\i}k
\end{center}

\begin{center}
Slovak Technical University, Bratislava, Slovakia
\end{center}

\bigskip

Abstract. The contribution provides a comparison of consequences stemming 
from D-brane theories and Expansive Nondecelerative Universe model, and 
calls attention to coincidence of the results arising from the mentioned 
approaches to a description of the Universe. It follows from the comparison 
that the effects of quantum gravitation should appear at the energy near to 
2 TeV.

\bigskip

D-brane theories of the Universe [1] are founded on the principle declaring 
that there are more than 3 space dimensions, and that the introduction of 
any further extra dimension limits its range. For two extra dimensions the 
related range is close to 1 mm, when dealing with seven extra dimensions, 
the range drops to $10^{ - 14}$m. It follows from the mathematical treatment 
of the issue by the D-brane theoretical approaches that some changes in the 
Newtonian gravitational law should occur on a scale below 1 mm. It is 
supposed that in the region $10^{ - 19}$m which corresponds to the energy of 
about 1 TeV, the effects of quantum gravitation should appear. 

Due to the Vaidya metric application [2], the model of Expansive 
Nondecelerative Universe (ENU) [2-5] enables to localize and quantify the 
energy of the gravitational field. For weak field conditions it adopts the 
form

\begin{equation}
\label{eq1}
\varepsilon _{g} = - {\frac{{R.c^{4}}}{{8\pi .G}}} = - 
{\frac{{3m.c^{2}}}{{4\pi .a.r^{2}}}}
\end{equation}

\noindent
where $\varepsilon _{g} $ is the density of gravitational field energy 
generated by a body with a mass $m$ at a distance $r$, $R$ is the scalar 
curvature and $a$ is the gauge factor being at present 

\begin{equation}
\label{eq:2a}
a \cong 1.3\times 10^{26} {\rm m}
\end{equation}
It is worth mentioning that the magnitude of $\varepsilon _{g} $ given by 
(\ref{eq1}) is closed to the $\theta _{0}^{0} $ component of the Einstein 
energy-momentum pseudotensor describing the density of gravitational energy 
also for strong field conditions [6]. In the ENU the critical density of 
gravitational energy is expressed as

\begin{equation}
\label{eq2}
\varepsilon _{crit} = {\frac{{3c^{4}}}{{8\pi .G.a^{2}}}}
\end{equation}

For the instances when

\begin{equation}
\label{eq3}
{\left| {\varepsilon _{g}}  \right|} = \varepsilon _{crit} 
\end{equation}

\noindent
it follows that the effective gravitational range $R_{ef} $ of a body having 
the gravitational radius $R_{g} $ is related to the gauge factor as

\begin{equation}
\label{eq4}
R_{ef} = \left( {R_{g} .a} \right)^{1 / 2}
\end{equation}

It follows of the above that at present there should exist a particle with 
the Compton wavelength $\lambda _{x} $ being equal to its effective 
gravitational range. It represents the lightest particle able to exert 
gravitational influence on its surrounding. The mass of the particle is 
expressed by

\begin{equation}
\label{eq5}
m_{x} = \left( {{\frac{{\hbar ^{2}}}{{G.a}}}} \right)^{1 / 3}
\end{equation}

At the time being
\begin{equation}
\label{eq:7a}
m_{x} \cong 10^{ - 28} \mathrm{kg} 
\end{equation}
and 
\begin{equation}
\label{eq:8a}
\lambda _{x} \cong 3.235\times 10^{ - 15}  \mathrm{m} 
\end{equation}
Based on the assumption that the ENU model is compatible with the 
superstring theory M [7], in which a number of space dimensions is 10, i.e. 
a number of extra dimensions is 7, we postulate that just relation (8) 
expresses their magnitude. In the case when

\begin{equation}
\label{eq6}
r \le \lambda _{x} 
\end{equation}

\noindent
in 10-dimensional space it must hold

\begin{equation}
\label{eq7}
E_{g} \approx r^{ - 8}
\end{equation}

At these conditions the gravitational energy attracting two particles with a 
mass $m$ may be expressed as

\begin{equation}
\label{eq8}
E_{g} = {\frac{{G.m^{2}.\lambda _{x}^{7}} }{{r^{8}}}}
\end{equation}

\noindent
and $r$ must be equal to the Compton wavelength of given particles,

\begin{equation}
\label{eq9}
r = {\frac{{\hbar} }{{m.c}}}
\end{equation}

In the limiting case, the energy $E_{g} $ will be identical to the rest 
energy $E_{o} $ of the particles, i.e.

\begin{equation}
\label{eq10}
E_{o} = {\frac{{\hbar .c}}{{r}}}
\end{equation}

It follows from the identity of (\ref{eq8}) and (\ref{eq10}) that

\begin{equation}
\label{eq11}
r^{9} = l_{Pc}^{2} .\lambda _{x}^{7} 
\end{equation}

\noindent
where the Planck length is
\begin{equation}
\label{eq:15a}
l_{Pc} = \left( {{\frac{{G.\hbar} }{{c^{3}}}}} \right)^{1 / 2} \cong 
1.614\times 10^{ - 35} \mathrm{m}
\end{equation}
It follows from~(\ref{eq:8a}), (\ref{eq11}) and~(\ref{eq:15a}) that
\begin{equation}
\label{eq:16a}
r \cong 9.66\times 10^{ - 20} \mathrm{m}
\end{equation}
which corresponds to the energy value
\begin{equation}
\label{eq:17a}
E \cong 2.0 \mathrm{TeV}
\end{equation}
We suppose that just at this energy, the effects of quantum gravitation will 
manifest themselves at present. 

A significant argument of our considerations lies in time evolution of
a number of extradimensions. At the initial stage of the Universe
expansion, all dimensions, including extradimensions, had to be of
Planck size which is in full agreement with relations (\ref{eq5}),
(\ref{eq:7a}), (\ref{eq:8a}), (\ref{eq11}), (\ref{eq:16a}), and
(\ref{eq:17a}) since they describe the phenomena dependent on the
cosmological time.

\subsection*{Conclusions}

\begin{enumerate}
\item The present contribution documents compatibility of predictions stemming 
from the ENU model and D-branes theory, particularly that related to the 
energy (2 TeV) at which the effects of quantum gravitation appears. 
\item Inevitability of time evolution of the size of extra dimensions and its 
rationalization is manifested. 
\end{enumerate}

\subsection*{References}

\begin{enumerate}

\item N. Arkani-Hamed, S. Dimopoulos, G.Dvali, Phys. Lett. B 429 (1998) 263 

\item P.C. Vaidya, Proc. Indian Acad. Sci., A33 (1951) 264

\item V. Skalsk\'y, M. S\'uken\'{\i}k, Astrophys. Space Sci. 178 (1991) 169

\item V. Skalsk\'y, M. S\'uken\'{\i}k, Astrophys. Space Sci. 181 (1991) 153

\item J. \v{S}ima, M. S\'uken\'{\i}k, Preprint gr-qc/9903090 

\item M. S\'uken\'{\i}k, J. \v{S}ima, Astrophys. Space Sci., submitted

\item M.B. Green, J.H. Schwarz, E. Witten, Superstring Theory, Vol. 1, 2. 
Cambridge University Press, Cambridge, 1989 
\end{enumerate}

\end{document}